\documentclass{article}
\usepackage{spconf,amsmath,graphicx,hyperref, amssymb,newunicodechar,adjustbox, booktabs}
\newunicodechar{（}{(}

\title{Subject-Invariant Cross-Modal Decoding of Perceived Speech from Brain Recordings}
\name{Aoke Zhang$^{1,2}$, Jing Chen$^{1,2,3}$}
\address{$^{1}$School of Intelligence Science and Technology, Peking University, Beijing, China\\ $^{2}$  Academy for Advanced Interdisciplinary Studies, Peking University, Beijing, China \\ $^{3}$ National Key Laboratory of General Artificial Intelligence, Beijing, China \\ janechenjing@pku.edu.cn}
\begin{document}
%
\maketitle
\begin{abstract}
Perceived speech decoding based on non-invasive brain-computer interface (BCI) signals has been extensively studied in recent years. Research in this field primarily faces two challenges: extracting neural representations with rich spatiotemporal information and achieving cross-subject generalization. Although separate studies have proposed methods to cope with these issues, a unified approach that simultaneously tackles both challenges remains lacking. To fill this gap, we propose the Subject-Invariant Cross-Modal Perceived Speech Decoding (SICMD) method, which integrates functional magnetic resonance imaging (fMRI) and magnetoencephalography (MEG). We conduct comprehensive analyses of the fusion method, fusion position, encoder architecture, and model inputs. Our results demonstrate that the proposed method improves Top-1, Top-10, and Rankacc by more than 10.6\%, 10.1\%, and 1.7\%, respectively, compared to baseline methods in cross-subject perceived speech decoding tasks, while reducing training costs by 88.8\% and 60.5\% compared to multi-subject and intra-subject decoding settings. Further visualization experiments also confirm the effectiveness of our approach.
\end{abstract}
\section{Introduction}
\label{sec:intro}

In recent years, perceived speech decoding has been extensively studied due to its applications in exploring neural mechanisms \cite{zou2026constituent,akbari2019towards} and assessment indicators in patients with aphasia \cite{Borrie_2021}. It also holds significant potential for advancing speech neuroprostheses \cite{wairagkar2025instantaneous, silva2024speech}, which can restore communication abilities for individuals with disabilities. Regarding neural signal acquisition devices, non-invasive techniques such as MEG and fMRI can avoid surgery and record whole-brain signals, making them popular research tools. However, non-invasive perceived speech decoding still faces two main challenges: extracting representations with rich spatiotemporal information and achieving cross-subject generalization.

Methods like Brainmagic \cite{defossez2023decoding} and ConvConcatNet \cite{wang2026hierarchical} align neural representations from MEG signals with corresponding speech features. However, due to the low spatial resolution of MEG, predicting high-level speech features such as semantics remains challenging, creating a bottleneck in decoding performance. Conversely, some studies leverage the high spatial resolution of fMRI to achieve text decoding \cite{chen2024open, lu2026brain}. Yet, because the temporal resolution of fMRI is typically on the order of seconds, capturing dynamic features is challenging, resulting in a high word error rate in synthesized text. To address these issues, fMRI-MEG Fusion (fMMF) framework \cite{zhang2025novel} integrates MEG and fMRI data to further improve decoding performance, achieving enhanced performance in perceived speech decoding. 

Although cross-modal methods can enhance performance, they are primarily designed for intra-subject decoding, utilizing neural data from the same subject for both training and evaluation. Due to significant individual differences among subjects \cite{giovannone2021individual,myers2024individual}, intra-subject decoding models exhibit poor generalizability to new target subjects. In contrast, cross-subject decoding achieves effective model transfer with high accuracy and efficiency by pre-training on source subjects, followed by fine-tuning and evaluation on the target subject \cite{zhang2026cross}. A common challenge in both cross-modal and cross-subject decoding is how to effectively extract subject-consistent information, as the consistency in responses to the same stimuli during the experiment likely reflects task-related information \cite{nastase2019measuring,hasson2004intersubject}. Moreover, extracting subject-consistent information plays a crucial role in enhancing cross-subject transferability. Cross-Subject Perceived Speech Decoding (CPSD) framework \cite{zhang2026cross} tackles this issue primarily by remapping data from different subjects into a reference space, achieving consistent performance improvements. However, CPSD relies exclusively on MEG/EEG data, which limits its ability to capture features with rich spatiotemporal detail.

\begin{figure*}[t]
\centering
\includegraphics[width=18cm, height=5.84cm]{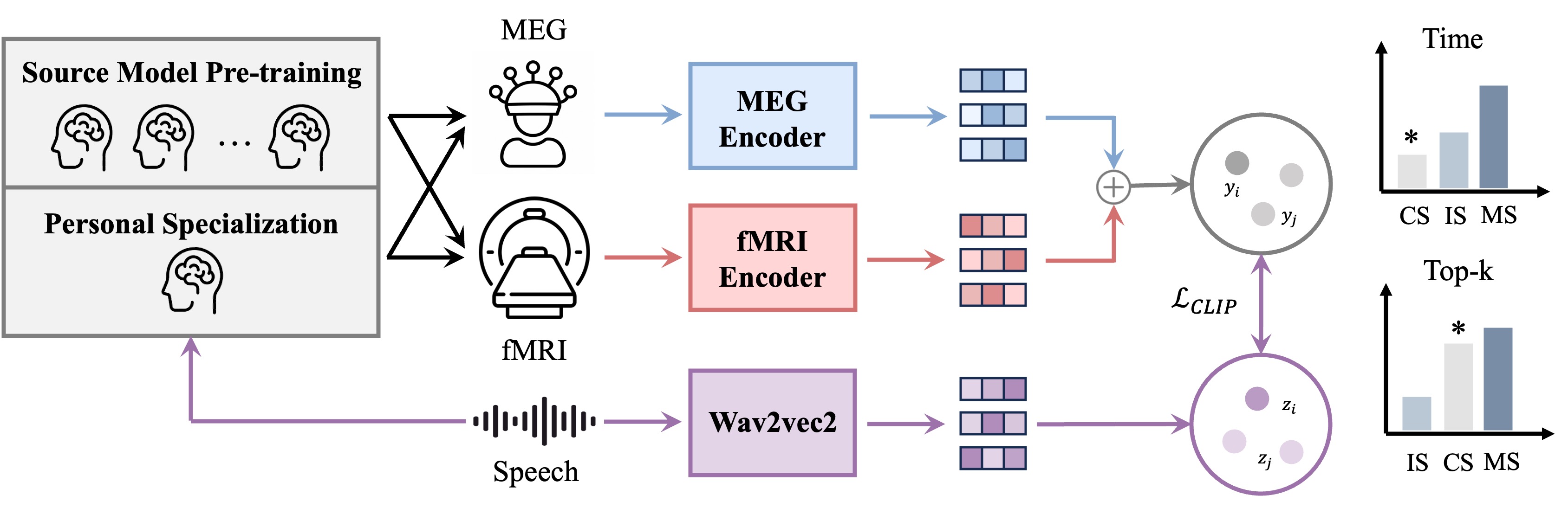}
\caption{Framework of SICMD. MS, IS, and CS refer to Multi-Subject, Intra-Subject, and Cross-Subject, respectively.}
\label{CPSD}
\end{figure*}

To effectively combine cross-modal and cross-subject methods, we propose SICMD, a model that learns neural representations enriched with spatiotemporal information and exhibits strong transferability to target subjects. To thoroughly investigate this problem,  we provide a comprehensive analysis of the fusion method, fusion position, encoder architecture, and model inputs. Additionally, further visualization analysis is presented to demonstrate the effectiveness of our proposed approach.

\section{Methods}
\label{sec:methods}

\subsection{Model Overview}

MEG encoder of SICMD consists of the following components: the Positional Encoding-based Spatial Attention (PESA) module \cite{zhang2026cross}, a $1\times 1$ CNN layer, the subject layer \cite{defossez2023decoding}, and a brain encoder. The fMRI encoder is composed simply of linear encoding modules.

\subsubsection{PESA}

To explicitly extract subject-consistent information, thereby enhancing cross-subject transferability and optimizing model training, PESA \cite{zhang2026cross} remaps MEG signals from different subjects into a standardized reference space. Let $T$ denote the time dimension, $C$ the channel dimension, and $D$ the hidden dimension, which is set to 270 in this experiment. The standardized reference space $P=(r_{n\delta,i})_{n=1,...,D}^{i=1,...,T}$ is first established using positional encoding by stacking embeddings selected at equal intervals of $\delta=15$ from \eqref{pe}: 

\begin{equation}
r_{ni}= \label{pe}
\begin{cases}
    \text{sin}(\frac{n}{L^{i/D}})\text{, \quad if i is even,} \\
    \text{cos}(\frac{n}{L^{(i-1)/D}})\text{, if i is odd.}
\end{cases}
\end{equation}

\noindent where $L$ is a user-defined scalar and is set to 10000 as convention. On the other side, sensor locations $X_{location} \in \mathbb{R}^{C\times 2}$ extracted by MNE-Python function find\_layout are first processed by an MLP module consists of three blocks \eqref{mlp}, each including a linear layer, layer normalization and GELU activation function.  

\begin{equation}
    Y = \text{MLP}(X_{location}), Y\in\mathbb{R}^{C\times T} \label{mlp}
\end{equation}

\noindent To measure the contribution of each channel to different locations in the standardized reference space, similarity between the MLP output $Y$ and position embeddings $P$ is computed, as shown in \eqref{W}:

\begin{equation}
    W = \text{Softmax}(YP^{T}), W\in\mathbb{R}^{C\times D} \label{W}
\end{equation}

\noindent The similarity matrix $W$ is then applied to the input data $X\in \mathbb{R}^{C\times T}$ to get the hidden output $H$ as in \eqref{final}:

\begin{equation}
    H = W^{T}X,H\in\mathbb{R}^{D\times T}\label{final}
\end{equation}

\subsubsection{Brain Encoder}
ConvBlocks from ConvConcatNet \cite{wang2026hierarchical} are employed as part of the MEG encoder in the SICMD method. This architecture enables the extraction and fusion of neural response patterns at multiple hierarchical levels through iterative processing and has been applied in perceived speech decoding tasks. The primary computational process of this encoder is as follows:

\begin{equation}
    H_{i}=
    \begin{cases}
        ConvBlock(Concat(H_{0},\mathbf{0}), i=1 \\
        ConvBlock(Concat(H_{0}, H_{i-1}),i\geq 2
    \end{cases}
\end{equation}

\noindent where $H_{i}$ denotes the output from the $i$-th ConvBlock. In SICMD, we have maintained the design and parameters used in the original model, as detailed in \cite{wang2026hierarchical}. The fMRI encoder is implemented as a simple MLP consisting of three encoding modules followed by a linear layer. Each encoding module includes a linear layer, a batch normalization layer, a ReLU activation function, and dropout regularization with a probability of 0.3. The input dimension of the first linear layer matches the fMRI data dimension, while the subsequent layers have a dimension of 1024.

\subsection{Training Pipeline}

A leave-one-subject-out approach is used to comprehensively evaluate our proposed method. We iteratively select each subject in the dataset as the target subject, with the remaining subjects serving as source subjects. The training pipeline of the SICMD method is illustrated in Figure \ref{CPSD}.

\subsubsection{Source Model Pre-training}

SICMD model requires three types of input during the source model pre-training stage: MEG segments, sensor locations, and subject IDs. First, the MEG segments are remapped by the PESA module using the sensor locations, then passed through a $1 \times 1$ convolutional layer. The resulting hidden outputs, along with the subject IDs, are processed by the subject layer. The hidden outputs from the subject layer are then fed into the MEG brain encoder. Finally, the model outputs are aligned with the corresponding wav2vec2 representations.

\subsubsection{Personal Specialization}

Before fine-tuning the model, subject layer for the target subject needs to be initialized using the method from \cite{zhang2026cross}. Subject consistency is utilized using CorrCA algorithm \cite{dmochowski2012correlated, parra2018correlated} to extract consistent component from source subject layers \eqref{corrca}:

\begin{equation}
    \label{corrca}
    \hat{w}=\mathop{\arg\max}\limits_{w} \frac{w^{T}X_{1}X_{2}^{T}w}{\Vert X_{1}^{T}w \Vert \Vert X_{2}^{T}w \Vert} 
\end{equation}

\noindent where $X_{1}$, $X_{2}$ refer to input data matrices and $w$ refers to the weight matrix, which all belong to $\mathbb{R}^{D\times D}$. Next, consistent components of source subject layers are averaged through \eqref{average} to obtain the initialized subject layer for the target subject.

\begin{equation}
    \label{average}
    X_{I} = \frac{1}{I-1}\sum_{i=1}^{I-1}X_i^{T}\hat{w}
\end{equation}

\noindent where $I$ refers to index of the target subject, and $\{1,...,I-1\}$ refer to the indices of the source subjects. After initialization, neural representations extracted from the model are realigned with the corresponding speech representations.

\subsection{Modality Fusion}
\label{sec:mf}

To achieve modality fusion of asynchronous data, SICMD employs the fusion method from fMMF \cite{zhang2025novel}. This approach leverages differences in the rate of information change across various speech levels and the predictive capabilities of fMRI and MEG for distinct levels of speech information.

\section{Experiments}
\label{sec:experiments}

\subsection{Dataset \& Preprocessing}

The performance of SICMD was evaluated using a publicly available multimodal neuroimaging dataset \cite{wang2022synchronized}. This dataset includes fMRI and MEG recordings from 12 subjects listening to the same continuous Chinese speech across 60 trials. The fMRI signals were acquired with a repetition time (TR) of 0.71 seconds. The MEG signals were recorded using a 306-channel system at a sampling rate of 1000 Hz. The preprocessing method described in \cite{zhang2025novel} was retained in this study.

\subsection{Implementation Details}
We divided the dataset into 70\% training set, 10\% validation set, and 20\% test set based on trial indices, ensuring no overlap between data from different trials. CLIP loss \cite{radford2021learning} was used as the loss function during training. The batch size and test sample size for the experiments were set to 128. The Adam optimizer, with a learning rate of $3 \times 10^{-4}$, was employed to update the model parameters. Early stopping was applied with a patience of 10 epochs. Top-k accuracy ($k=1,10$) and Rankacc were used to evaluate performance in perceived speech decoding, ensuring comparability with fMMF \cite{zhang2025novel} and CPSD \cite{zhang2026cross}. All experiments were conducted on a single NVIDIA RTX 3090 GPU.

\subsection{Experimental Results}

\subsubsection{Perceived Speech Decoding}

\begin{table}[h]
\caption{Perceived speech decoding results.}
\begin{adjustbox}{width=\columnwidth,center}
\centering
\begin{tabular}{l|l|l|l}
\toprule
{} & {Top-1(\%)} & {Top-10(\%)} & {Rankacc(\%)}\\
\midrule
Random & 0.1 $\pm$ 0.0 & 7.8 $\pm$ 0.0 & 50.0 $\pm$ 0.0 \\
\midrule
fMRI Encoder & 1.5 $\pm$ 0.4 & 6.2 $\pm$ 0.8 & 55.2 $\pm$ 1.5 \\
Brainmagic & 22.2 $\pm$ 11.8 & 59.8 $\pm$ 17.7 & 83.9 $\pm$ 7.6 \\
ConvConcatNet & 26.0 $\pm$ 9.4 & 68.9 $\pm$ 12.1 & 84.7 $\pm$ 6.4 \\
\midrule
CPSD & 30.8 $\pm$ 12.9 & 71.9 $\pm$ 14.9 & 89.2 $\pm$ 8.4 \\
fMMF & 32.9 $\pm$ 14.5 & 72.6 $\pm$ 16.1 & 91.7 $\pm$ 5.9 \\
\midrule
\textbf{SICMD} & \textbf{43.5} $\pm$ \textbf{13.7} & \textbf{82.7} $\pm$ \textbf{12.7} & \textbf{93.4} $\pm$ \textbf{5.0} \\
\bottomrule
\end{tabular}
\end{adjustbox}
\label{main}
\end{table}

The averaged perceived speech decoding results across subjects are shown in Table \ref{main}, where Random denotes the chance level for decoding. The results of intra-subject decoding using only MEG or fMRI data are presented in the second part of Table \ref{main}. Compared to fMRI Encoder, MEG-based models demonstrate significantly higher decoding performance, highlighting the importance of the dynamic features of neural signals. Among these models, ConvConcatNet achieves a Top-10 accuracy that is 9.1\% higher than Brainmagic in Top-10 accuracy and is therefore selected as the brain encoder in the MEG pipeline. The cross-subject decoding method CPSD \cite{zhang2026cross} and the cross-modal decoding method fMMF \cite{zhang2025novel} outperform ConvConcatNet by over 3\% and 3.7\%, respectively. By unifying cross-modal and cross-subject decoding, SICMD achieves the highest decoding performance of 43.5\%, 82.7\%, and 93.4\% in Top-1, Top-10, and Rankacc metrics, respectively—exceeding baseline methods by 10.6\%, 10.1\%, and 1.7\%. All performance improvements of SICMD shown in Table \ref{main} were evaluated using pairwise t-tests and are statistically significant ($p < 0.001$). 

Training time was also compared to evaluate the efficiency of the SICMD model. The average number of training steps for the multi-subject, intra-subject, and personal specialization stages were 5,595.5, 1,593.4, and 629.3, respectively. The personal specialization stage required 88.8\% and 60.5\% fewer training steps compared to the multi-subject and intra-subject stages. This demonstrates that our method avoids retraining through multi-subject decoding and effectively leverages the prior information obtained during the source model pre-training stage.

\subsubsection{Ablation Study}
\label{sec:ab}

Ablation studies were conducted to evaluate the rationale behind our design choices. Comparisons of different modality fusion methods in SICMD are presented in Table \ref{fm}. The results demonstrate that the fusion method described in Section \ref{sec:mf} yields the highest decoding performance and shows statistical significance ($p<0.001$) compared to the Random, Cross-Attention, and Multiply methods. 

\begin{table}[h]
\caption{Ablation study of the fusion methods.}
\begin{adjustbox}{width=\columnwidth,center}
\centering
\begin{tabular}{l|l|l|l}
\toprule
{} & {Top-1(\%)} & {Top-10(\%)} & {Rankacc(\%)}\\
\midrule
Random & 0.1 $\pm$ 0.0 & 7.8 $\pm$ 0.0 & 50.0 $\pm$ 0.0 \\
\midrule
Cross-Attention & 1.1 $\pm$ 0.4 & 9.0 $\pm$ 1.2 & 57.6 $\pm$ 2.2\\
Multiply & 31.8 $\pm$ 14.8 & 71.5 $\pm$ 16.0 & 92.1 $\pm$ 4.4\\
Concatenate & 39.4 $\pm$ 16.8 & 78.6 $\pm$ 15.0 & 87.7 $\pm$ 6.1\\
\textbf{SICMD} & \textbf{43.5} $\pm$ \textbf{13.7} & \textbf{82.7} $\pm$ \textbf{12.7} & \textbf{93.4} $\pm$ \textbf{5.0}\\
\bottomrule
\end{tabular}
\label{fm}
\end{adjustbox}
\end{table}

In Table \ref{position}, we compare the results of modality fusion across different module outputs, including PESA, CNN, the subject layer, and the brain encoder. The results indicate that applying modality fusion to the outputs from brain encoder leads to significant performance improvements ($p<0.001$).

\begin{table}[h]
\caption{Ablation study of the fusion positions.}
\begin{adjustbox}{width=\columnwidth,center}
\centering
\begin{tabular}{l|l|l|l}
\toprule
{} & {Top-1(\%)} & {Top-10(\%)} & {Rankacc(\%)}\\
\midrule
Random & 0.1 $\pm$ 0.0 & 7.8 $\pm$ 0.0 & 50.0 $\pm$ 0.0 \\
\midrule
PESA & 36.3 $\pm$ 12.4 & 76.3 $\pm$ 13.2 & 91.6 $\pm$ 6.6 \\
CNN & 32.0 $\pm$ 15.0 & 71.3 $\pm$ 15.6 & 89.8 $\pm$ 7.7 \\
Subject Layer & 35.0 $\pm$ 15.4 & 73.2 $\pm$ 16.8 & 91.3 $\pm$ 7.6 \\
\textbf{SICMD} & \textbf{43.5} $\pm$ \textbf{13.7} & \textbf{82.7} $\pm$ \textbf{12.7} & \textbf{93.4} $\pm$ \textbf{5.0} \\
\bottomrule
\end{tabular}
\label{position}
\end{adjustbox}
\end{table}

\begin{figure}[ht]
\centering
\includegraphics[width=8.8cm, height=3.6cm]{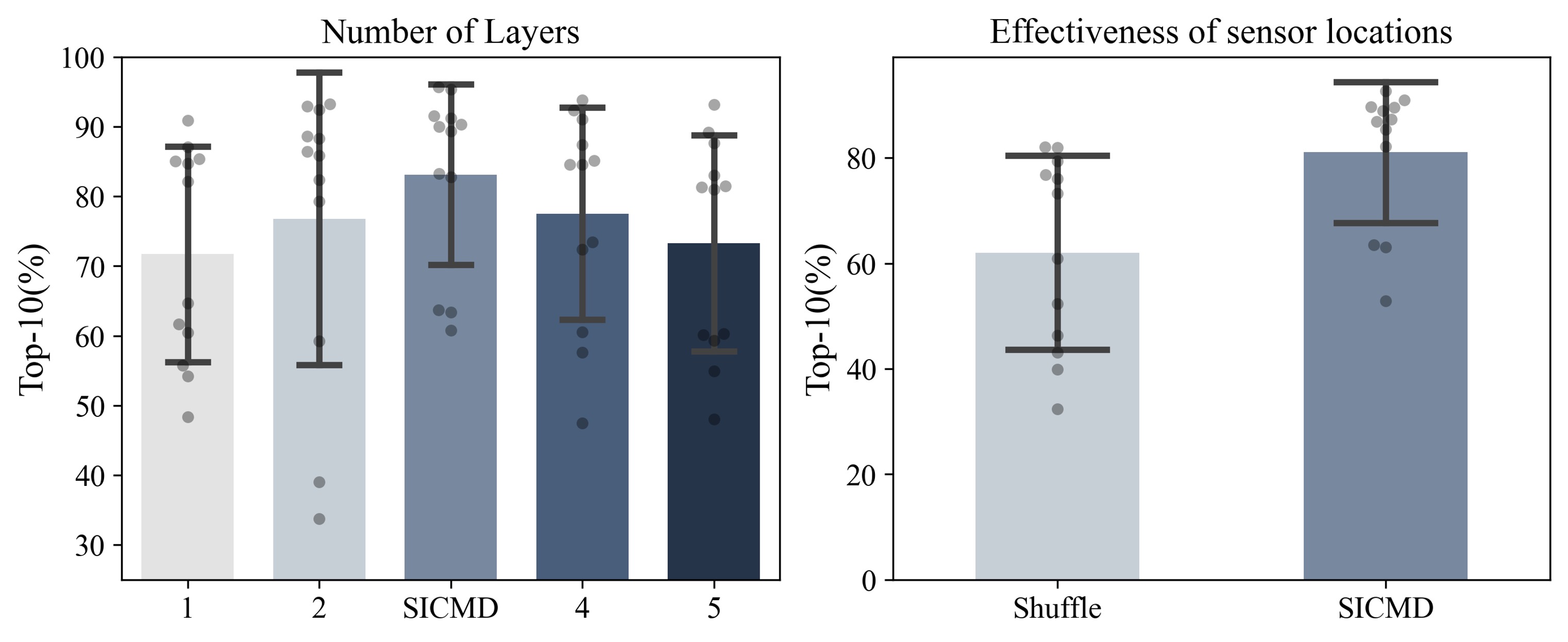}
\caption{Ablation study of model architecture and input.}
\label{ablation-3}
\end{figure}

An ablation study on the number of layers in the fMRI Encoder is presented in the left side of Figure \ref{ablation-3}. The results demonstrate the effectiveness of the hyperparameters used in SICMD. To verify the importance of sensor locations as input, we randomly shuffled the order of sensor locations and retrained the model, as shown on the right side of Figure \ref{ablation-3}. The Top-10 accuracy under the shuffled condition decreased by 19.0\% ($p<0.001$).

\subsubsection{PESA Analysis}

\begin{figure}[ht]
\centering
\includegraphics[width=8.59cm, height=4cm]{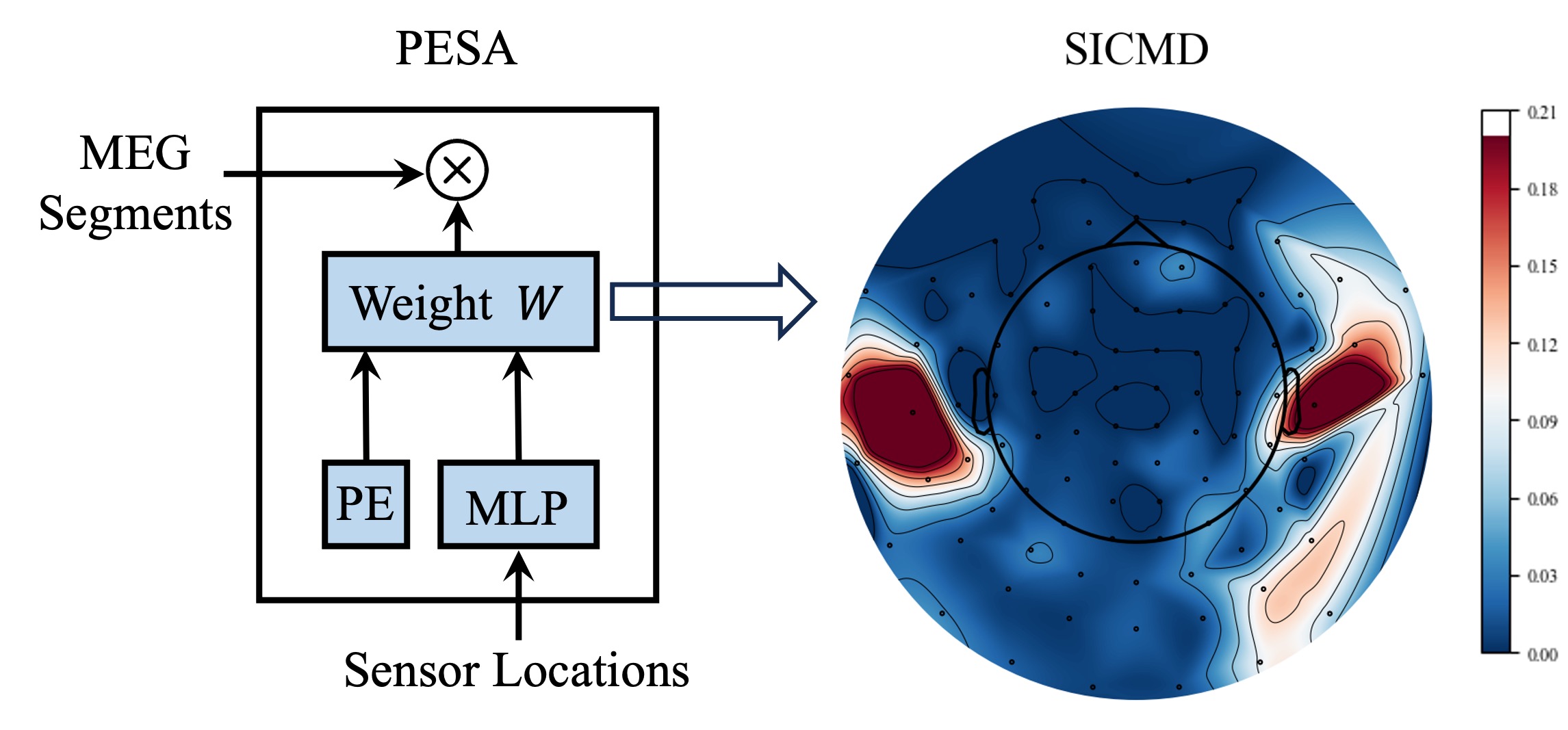}
\caption{Visualization of channel weights in PESA.}
\label{topo}
\end{figure}

We further visualized the channel weights using the method described in \cite{zhang2026cross}. The results shown in Figure \ref{topo} indicate that the weights near the bilateral temporal lobes were higher, demonstrating that the model automatically learned effective information.

\section{Conclusion}
\label{sec:conclusion}

In this research, we propose a subject-invariant cross-modal perceived speech decoding method, SICMD. This model leverages the fusion of MEG and fMRI modalities to obtain representations rich in spatiotemporal information and extracts consistent information across subjects to simultaneously enhance transferability and performance. Extensive analyses including training time comparison, ablation studies, and visualization of weight matrix in PESA demonstrate the effectiveness and efficiency of our proposed method. In future work, we aim to extend existing methods to EEG-fNIRS systems with improved applicability.


\section{Acknowledgment}

This work was supported by the STI 2030—Major Projects (No. 2021ZD0201500), the Hubei Provincial Key R\&D Program of Technological Innovation (JSCX202601690), the High-performance Computing Platform of Peking University and the Biomedical Computing Platform of National Biomedical Imaging Center, Peking University.

\bibliographystyle{IEEEbib}
\bibliography{strings,refs}

\end{document}